\documentclass[twocolumn,pra]{revtex4}
\usepackage[koi8-r]{inputenc}
\usepackage[english,russian]{babel}
\usepackage{amsmath}
\usepackage{amssymb}
\usepackage{epsfig}

\begin{document}
\title{Нелинейная динамика волн над неоднородно периодическим дном}

\author{В.\,П.\,Рубан}
\email{ruban@itp.ac.ru}
\affiliation{Институт теоретической физики им.~Л.\,Д.\,Ландау РАН, 142432, 
Черноголовка, Россия}

\date{\today}

\begin{abstract}
Путем численного моделирования точных уравнений движения (в конформных
переменных) для плоских нестационарных потенциальных течений идеальной
жидкости со свободной поверхностью над сильно неоднородным профилем дна
обнаружен эффект нелинейного сжатия длинного волнового пакета при его
Брэгговском отражении от участка с плавно нарастающей высотой периодически
расположенных барьеров. При этом образуется короткий и высокий пакет стоячих
волн с резкими гребнями, который затем трансформируется в обратную волну.
Существенно, что по частоте падающей волны эффект максимален не в середине
обусловленной барьерами спектральной щели, а ближе к ее верхнему краю, когда
прямая волна успевает проникнуть достаточно далеко в рассеивающую область
и там, вместе с возникшей обратной волной, сформировать на некоторое время
подобие Брэгговского солитона.

\vspace{1mm}

\end{abstract}

\maketitle

{\bf Введение.}
Волны на свободной поверхности жидкости могут сильно взаимодействовать с
неоднородностями топографии дна, и это оказывает влияние на протекание
многих природных и лабораторных процессов. В частности, большой интерес
представляет многократное рассеяние волны над периодически либо случайно
расположенными препятствиями (см., например,
\cite{DH1984,HM1987,Y2003,CLK2004,HKY2005,R2004,R2008-a,R2008-b,R2012,
CGCA2015,LLZ2016,ZB2019,RNF2024,RF2024} и ссылки там).
Если длина волны $\lambda=2\pi/\kappa$ примерно в два раза превышает локальный
пространственный период донной неоднородности $\Lambda$ (условие Брэгговского
резонанса), то возникает линейная связь между прямой и обратной волной,
а в частотном спектре открывается запрещенная зона (щель). Поэтому массив из
нескольких барьеров на дне канала становится практически непреодолимой
преградой для волны с несущей частотой внутри щели.

С другой стороны, нелинейность гравитационных волн на достаточно глубокой
воде устроена таким образом, что над периодическим дном теоретически возможны
так называетые Брэгговские (или щелевые) солитоны --- долгоживущие уединенные
структуры, представляющие собой (неподвижные в простейшем случае) пакеты
стоячих волн. Они соответствуют стационарным решениям следующей приближенной
системы уравнений для огибающих $a_\pm(x,t)$ главных гармоник прямой (бегущей
вправо) волны и обратной (бегущей влево) волны \cite{R2008-b}:
\begin{equation}
i\Big(\frac{\partial_t}{\omega_*}\pm \frac{\partial_x}{2\kappa}\Big)a_\pm=
\tilde\Delta a_\mp+\frac{1}{2}(|a_\pm|^2-2|a_\mp|^2)a_\pm,
\label{a_pm}
\end{equation}
где $\omega_*=\sqrt{g\kappa}$ --- частота волны на бесконечно глубокой воде
($g=9.81$ м/c$^2$ --- ускорение свободного падения),
а малый действительный параметр $\tilde\Delta$ определяет полуширину щели
$\Delta=\omega_*\tilde\Delta$.
При этом главная гармоника вертикального отклонения свободной поверхности
дается формулой
\begin{eqnarray}
\eta(x,t)&=&\frac{1}{\kappa}\mbox{Re}[
 a_+(x,t)e^{ i\kappa x-i\omega_*(1-\epsilon)t}\nonumber\\
&&\qquad+a_-(x,t)e^{-i\kappa x-i\omega_*(1-\epsilon)t}],
\end{eqnarray}
где малый параметр $\epsilon=\exp(-2\kappa h_0)>\tilde\Delta$ учитывает конечную
эффективную (конформную) глубину слоя жидкости \cite{R2004,R2008-b}. Щелевые
солитоны наблюдались в высокоточных численных экспериментах \cite{R2008-b},
где начальным состоянием была горизонтальная поверхность жидкости и задавалось
соответствующее поле скорости. Понятно, что такой способ одномоментного
возбуждения волны не может быть реализован обычным волнопродуктором.
Поэтому до сих пор нет экспериментального подтверждения существования
подобных солитонов.

Естественно приходит мысль создать бегущую волну на некотором отдалении
от массива барьеров и посмотреть, не сформируется ли в процессе рассеяния
что-либо похожее на Брэгговский солитон. Но для начала необходимо провести
соответствующее численное моделирование, причем с полным учетом нелинейности
волн и сильной неоднородности дна. Надо сказать, что в такой постановке до
сих пор задача не решалась. Данная работа направлена на заполнение этого
пробела в знаниях. Здесь проведена серия численных экспериментов, в которых
на массив постепенно повышающихся барьеров набегал длинный волновой пакет
(около 50 волн) из области с горизонтальным дном либо из области с достаточно
большой глубиной, где на данной длине волны дно ``не чувствуется''.
Основным результатом этого исследования стало наблюдение того факта, что
при частотах падающей волны внутри запрещенной зоны, но ближе к ее верхнему
краю действительно происходит сильное сжатие волнового пакета (до 5-7 волн).
При этом амплитуда возникающей стоячей волны может увеличиваться в несколько
раз и достигать предельных значений, когда формируются резкие гребни.
Численный пример подобной концентрации волновой энергии приведен на рисунке 1.

{\bf Организация численных экспериментов.}
В работе моделировались плоские потенциальные течения идеальной несжимаемой
жидкости со свободной границей. Для таких течений разработан высокоточный
численный метод на основе использования так называемых конформных
переменных \cite{Ovs1,Ovs2}, которые представляют нестационарную область
течения как конформное отображение нижней полуплоскости --- для бесконечно
глубокой воды \cite{Z1,Z2,Z3,ChoiC1999}, либо горизонтальной полосы ---
при наличии непроницаемого, возможно пространственно неоднородного дна
\cite{R2004,R2012,Z4,R2005,R2008,R2020}. Форма свободной поверхности при
этом задается параметрически с использованием некоторой аналитической функции
$B(\zeta)$ (со свойством $B(\zeta+2\pi)=L_x+B(\zeta)$, где $L_x$ --- период
по горизонтальной координате):
\begin{eqnarray}
&&X^{(s)}(\vartheta,t)+iY^{(s)}(\vartheta,t)\nonumber\\ 
&&\qquad
=B(\vartheta+i\alpha(t)+[1+i\hat R_\alpha]\rho(\vartheta,t)-i\alpha_0),
\end{eqnarray}
где $\rho(\vartheta,t)$ --- неизвестная $2\pi$-периодическая по переменной
$\vartheta$ действительная функция двух переменных, $\alpha(t)$ --- неизвестная
действительная функция времени, $\hat R_\alpha$ --- диагональный
в Фурье-представлении оператор, умножающий $m$-ую гармонику $\rho_m(t)$ на
$i\tanh[m\alpha(t)]$. Принципиально важно, что комбинация 
$\vartheta+i\alpha(t)+[1+i\hat R_\alpha]\rho(\vartheta,t)$ является аналитическим
продолжением чисто действительной функции с действительной оси на горизонтальную
прямую $\vartheta+i\alpha(t)$. 

Замкнутые уравнения движения для неизвестных функций $\alpha(t)$, $\rho_m(t)$, 
а также для гармоник $\psi_m(t)$ поверхностного потенциала скорости приведены 
в работе \cite{R2004} (см. также их обобщение на случай подвижного дна 
\cite{R2005} и на случай постоянной завихренности \cite{R2008,R2020}). 
Интересно отметить, что эти точные уравнения оказались гораздо более 
компактными и удобными для численного моделирования двумерных течений над
неоднородным дном, нежели многие широко применяемые приближенные модели 
(см., например, 
\cite{Boussinesq2005,Boussinesq2006,Boussinesq2009,HOS2016}). Это удобство
и простота обусловлены двумя факторами: 1) наличием в современных языках
программирования библиотек элементарных аналитических функций комплексной
переменной; 2) высокоэффективной реализацией алгоритмов быстрого
преобразования Фурье. 

\begin{figure}
\begin{center} 
\epsfig{file=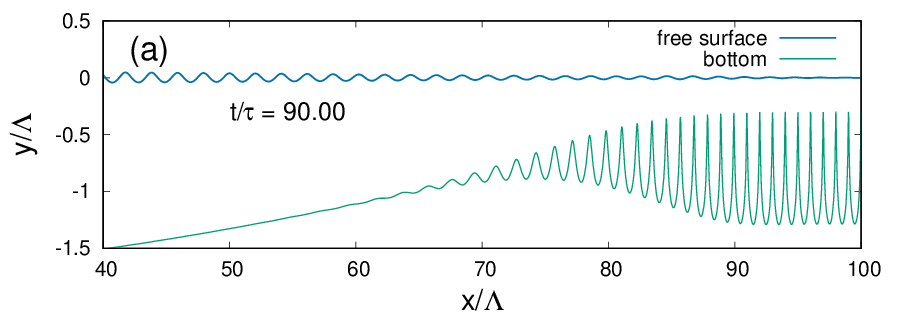, width=88mm}\\
\epsfig{file=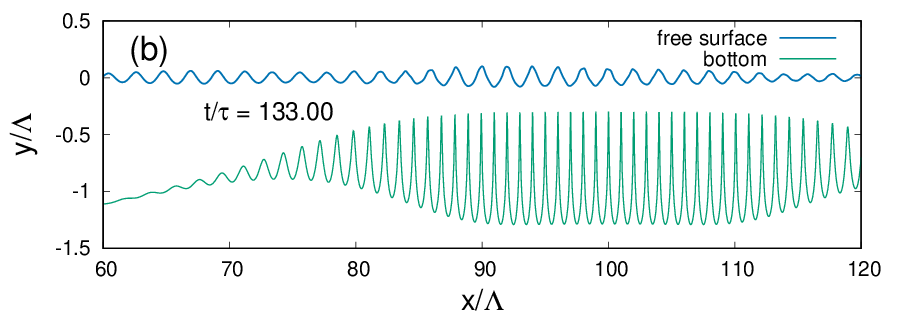, width=88mm}\\
\epsfig{file=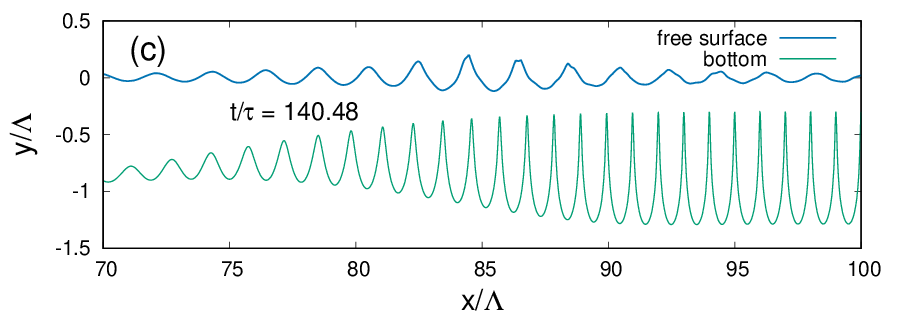, width=88mm}\\
\epsfig{file=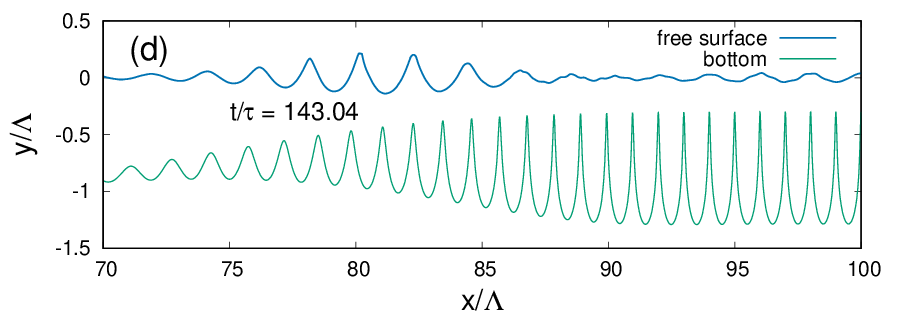, width=88mm}
\end{center}
\caption{Пример формирования высокой волны при рассеянии длинного 
волнового пакета на массиве локально периодических барьеров
(см. подробности в тексте).
}
\label{deep} 
\end{figure}

\begin{figure}
\begin{center} 
\epsfig{file=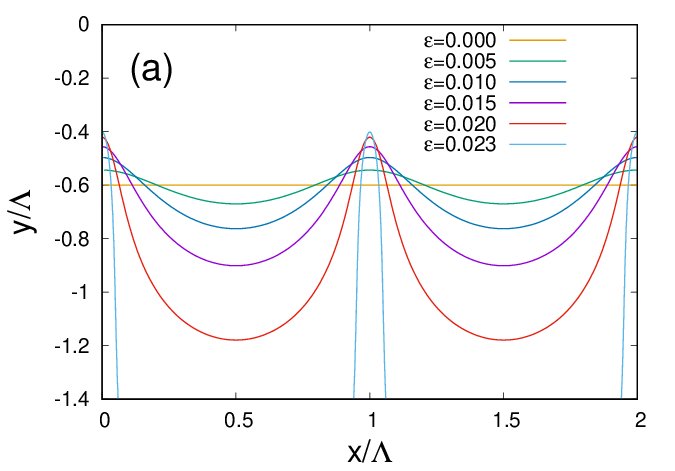, width=75mm}\\
\epsfig{file=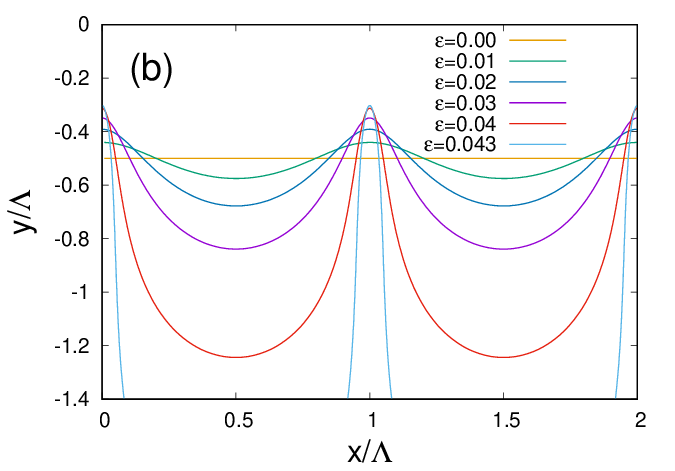, width=75mm}\\
\epsfig{file=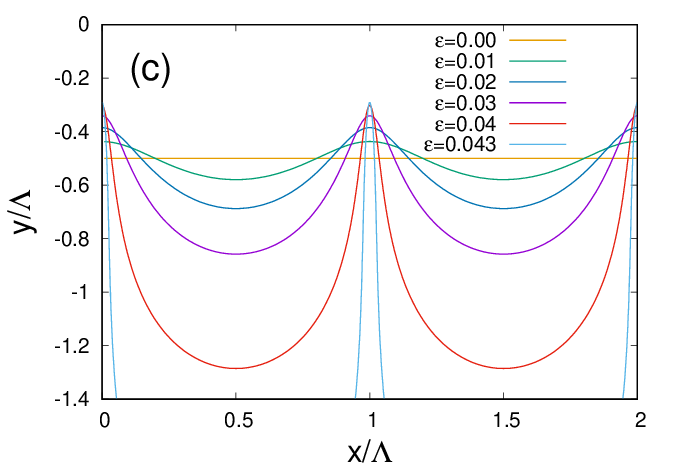, width=75mm}\\
\epsfig{file=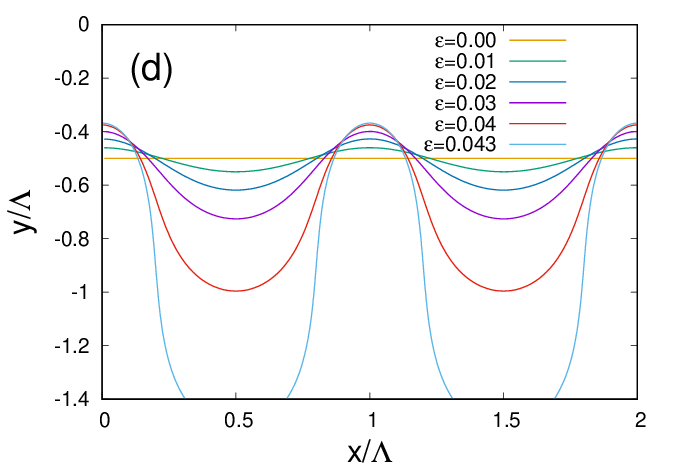, width=75mm}
\end{center}
\caption{Периодические профили дна при различных значениях параметров:
a) $\alpha_0/\Lambda_F=0.6$, $Q=0.90$; b) $\alpha_0/\Lambda_F=0.5$, $Q=0.90$;
c) $\alpha_0/\Lambda_F=0.5$, $Q=0.95$; d) $\alpha_0/\Lambda_F=0.5$, $Q=0.60$.
}
\label{bed_profiles} 
\end{figure}

Ключевым моментом при программировании является выбор аналитической функции 
$B(\zeta)$ и постоянного параметра $\alpha_0$, которые совместно определяют
неподвижный профиль дна по формуле
\begin{equation}
X^{(b)}(u)+iY^{(b)}(u)= B(u-i\alpha_0).
\end{equation}
Необходимо также позаботиться о том, чтобы выражение для производной $B'(\zeta)$
было задано самосогласованным образом.

Удобно подбирать функцию $B(\zeta)$ таким образом, чтобы при действительных
$\zeta$ она принимала чисто действительные значения. Тогда при $\alpha=\alpha_0$
и $\rho=0$ получается невозмущенная горизонтальная граница $y=0$. 

Если рассматривать строго периодические профили дна, то из физических соображений
следует, что имеется всего три наиболее существенных геометрических параметра:
расстояние от вершины барьера до линии $y=0$, высота барьера от впадины до вершины,
а также кривизна вершины. В данной работе за основу была взята следующая функция,
позволяющая (совместно с $\alpha_0$) регулировать указанные параметры:
\begin{equation}
F(\zeta, Q, N_b,\varepsilon)=\zeta+\frac{2iQ}{N_b}
\mbox{Ln}\,\frac{1+\varepsilon\exp(iN_b\zeta)}{1+\varepsilon\exp(-iN_b\zeta)},
\end{equation}
где $N_b$ --- целое число барьеров на периоде $2\pi$, так что безразмерный период
неоднородности $\Lambda_F=2\pi/N_b$;  малый параметр $\varepsilon\ll 1$ ``отвечает''
за высоту барьеров, а параметр $0<Q<1$ --- за скругление их вершин.
Примеры соответствующих профилей дна приведены на рисунке 2.
Видно, в частности, что малость параметра $\varepsilon$ вовсе не подразумевает
малую высоту барьеров. Более того, барьеры могут иметь даже бесконечную высоту
при критическом значении $\alpha_0=(\Lambda_F/2\pi)\ln(1/\varepsilon)$.
Структура глубоких впадин при этом не так важна, как высота и форма вершин,
а эффективная глубина определяется именно (конечным) параметром $\alpha_0$.
Поскольку поле скорости не проникает в глубокие впадины, их на практике можно
будет заменить горизонтальными участками, чтобы профиль дна не выглядел
слишком ``непривычным''.

Следует еще отметить, что по результатам наших численных экспериментов оптимальные
значения отношения $\alpha_0/\Lambda_F$ лежат в диапазоне примерно между 0.5 и 0.6.
Б\'{о}льшие глубины подразумевают слишком малые допустимые значения $\varepsilon$
и потому дают слишком малую ширину щели (параметр $\tilde \Delta =Q\varepsilon$
в главном приближении \cite{R2004,R2008-b}). Меньшие же значения
$\alpha_0/\Lambda_F\lesssim 0.4$ выводят волну из режима глубокой воды и
способствуют формированию свободных коротких волн, нежелательным образом
``засоряющих'' картину явления. Кроме того, особенности конформного отображения 
$F(\zeta)$ в верхней полуплоскости по принципу симметрии могут оказаться
недостаточно удаленными от вещественной оси и своим присутствием помешать
моделированию волн с высокими гребнями.

Как уже было сказано, нас интересуют профили дна с ровными участками и 
с плавно меняющейся высотой барьеров вдоль горизонтальной координаты $x$.
Этого можно добиться, положив величину $\varepsilon$ не константой, а 
относительно медленной аналитической функцией $\varepsilon(\zeta)$, с
действительными значениями на действительной оси. Боле того, для построения 
результирующей функции $B(\zeta)$ можно использовать дополнительное конформное
отображение $G(F)$, так что
\begin{equation}
B(\zeta)=G(F(\zeta, Q, N_b,\varepsilon(\zeta))),
\end{equation}
где аналитическая функция $G(F)$ также принимает действительные значения
на действительной оси. Это отображение вносит неоднородность эффективной 
глубины и шага барьеров, поскольку в главном приближении локально 
``растягивает'' относительно малое расстояние $\Lambda_F$ между соседними
барьерами и все их размеры фактором $|G'(F)|$. Например, рисунок 1 соответствует
выбору параметров $N_b=100$, $\alpha_0/\Lambda_F=0.5$, $Q=0.95$, а также функциям
\begin{equation}
\varepsilon(\zeta)=0.04\exp(-2.0[1+\cos\zeta]^3),
\end{equation}
\begin{equation}
G(F)=F+0.66\sin(F)+0.08\sin(2F).
\label{GF}
\end{equation}
Расстояние между соседними барьерами в таком случае является медленной 
функцией $\Lambda(x)\approx 2\pi/200$ --- в меру неоднородности производной 
$G'(F)\approx 0.5$ вблизи $\zeta=\pi$.

В уравнении Бернулли безразмерное ускорение свободного падения полагалось
равным $\tilde g=1$, а длина вычислительной области по координате $x$ была равна
$2\pi$. В результате такого выбора единица безразмерного времени соответствует
интервалу $\tau=\sqrt{L_x/2\pi g}$, что дает 1.8 с при длине бассейна $L_x=200$ м.

Теперь необходимо сказать о способе возбуждения бегущего волнового пакета. 
В наших численных экcпериментах волна создавалась локализованным во времени 
и пространстве воздействием давления $P(x,t)$ на изначально горизонтальную
неподвижную свободную границу жидкости, что сводилось к добавлению
соответствующего члена в динамическое граничное условие (уравнение Бернулли).
Внешнее воздействие работало в течение периода времени $0\leq t\leq T_f$ по закону
\begin{eqnarray}
P(x,t)&=&6.75 A_f(t/T_f)^2(1-t/T_f)\cos(k_f x-\omega_f t)\nonumber\\
&&\times\exp(-10[1-\cos(x+0.5\pi)]),
\end{eqnarray}
где амплитуда $A_f\sim 3.0\times 10^{-5}$ определяла силу воздействия,
$k_f$ --- волновое число (вблизи резонансного значения $k_{\rm res}=100$ при
$\Lambda=2\pi/200$), а частота создаваемой волны
$\omega_f=\sqrt{k_f\tanh(\alpha_0 k_f)}$ в случае $G(F)=F$ (нет глубоких участков),
и $\omega_f=\sqrt{k_f}$ в случае $G(F)$, заданной выражением (\ref{GF}),
когда волна набегает с большой глубины. Время воздействия было $T_f=90.0$
в представленных примерах. Это создавало длинный волновой пакет с характерной
амплитудой около $0.1\Lambda$ от гребня волны до ее впадины, который в момент
времени $T_f$ только начинал взаимодействовать с барьерами своим передним краем
[как показано на рисунке 1(a)].
При $t>T_f$ накопленная механическая энергия системы сохранялась с точностью до
семи-восьми десятичных знаков на протяжении многих десятков временных единиц,
что говорит об очень высоком качестве моделирования.

{\bf Результаты моделирования.}
В первой серии проведенных численных экспериментов был использован описанный
выше профиль дна с глубоким участком. Характерный пример взаимодействия волны
с массивом барьеров представлен на рисунке 1, который соответствует параметрам
$A_f=2.4\times 10^{-5}$, $k_f=95$. В качественном согласии с упрощенной
моделью (\ref{a_pm}) прямая волна проникала в область над барьерами, после
чего создавала там обратную волну. Обе волны сосуществовали некоторое время в
форме стоячей волны умеренной амплитуды [как показано на рисунке 1(b)].
Затем нелинейность вкупе с наведенной линейной дисперсией формировали
короткую и высокую группу почти стоячих волн с резкими гребнями. Размах
колебаний поверхности достигал при этом величины около $0.4\Lambda$,
что близко к эффективной глубине $0.5\Lambda$. Эта группа затем
относительно медленно передвигалась в обратном направлении, как это
видно из сравнения рисунков 1(c) и 1(d). В данном конкретном примере симуляция
была вынужденно остановлена в некоторый момент времени, так как даже большого
числа $N=2^{20}$ точек дискретизации по переменной $\vartheta$ оказалось
недостаточно для адекватного моделирования слишком резких гребней. В других
случаях, с меньшими значениями $A_f<2.1\times 10^{-5}$, до подобной 
(квази-)сингулярности дело не доходило и удавалось проследить за обратным
движением высокой группы и ее превращением в лево-бегущую волну после выхода
на глубокий участок.

Для лучшего понимания зависимости динамики от частоты падающей волны была 
проведена также вторая серия численных экспериментов, где дополнительное
отображение $G(F)$ не применялось, $N_b=200$, $\alpha_0/\Lambda_F=0.5$, $Q=0.90$,
а функция $\varepsilon(\zeta)$ имела вид
\begin{equation}
\varepsilon(\zeta)=0.02[1+\tanh(-2\cos(\zeta)+0.5\cos(2\zeta)-0.5)].
\label{tanh}
\end{equation}
В отличие от предыдущего примера, здесь расстояние между соседними барьерами
оставалось одинаковым на всем протяжении области неоднородности.

Было проведено моделирование с параметрами $k_f= 96, 100, 102, 103, 104, 105$.
Весьма примечательно, что резонансное значение $k_f=100$, при котором частота
волны находится в середине запрещенной зоны, не дало ничего похожего на высокую
и короткую волновую группу. Волна отразилась от массива препятствий, почти не
проникая в него, наподобие отражения от вертикальной стенки. Зато в случаях
$k_f= 103, 104, 105$ сжатие пакета и формирование высоких и резких гребней
происходило по уже знакомому нам сценарию. Наиболее выразительный  случай с
$k_f= 104$, $A_f=2.5\times 10^{-5}$ проиллюстрирован на рисунке 3. Нижняя кривая
там (выглядящая из-за разности вертикального и горизонтального масштабов как
гребенка с плавно неоднородной высотой зубьев) --- профиль дна. Профили
свободной поверхности представлены с интервалом по времени $1.28\tau$, что
примерно соответствует двум периодам волны. Все волновые профили, кроме самого 
последнего, ``раздвинуты'' в вертикальном направлении, чтобы избежать наложения.
Видно, что сжатие волнового пакета произошло здесь довольно аккуратным образом.

\begin{figure}
\begin{center} 
\epsfig{file=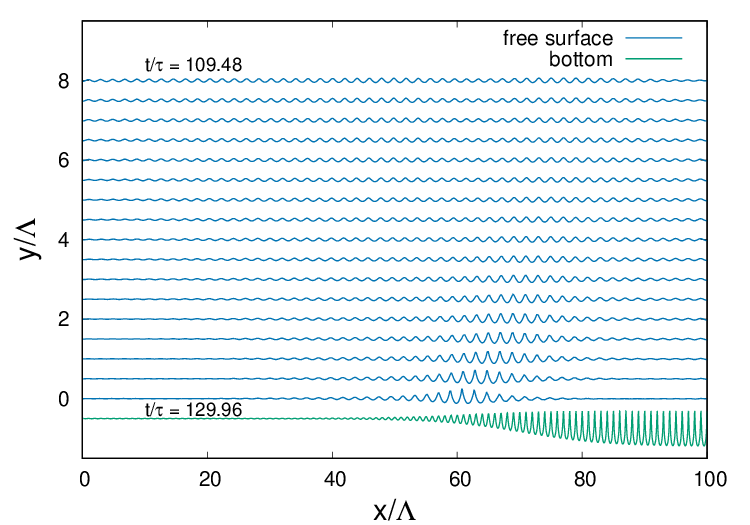, width=90mm}
\end{center}
\caption{Формирование высокой группы стоячих волн из набегающей волны
при однородной эффективной глубине.
}
\label{flat} 
\end{figure}

Были также проведены эксперименты с более резким переходом от плоского дна
к барьерам [с б\'{о}льшим в 2 либо в 4 раза общим коэффициентом под знаком
гиперболического тангенса в выражении (\ref{tanh})]. Формирование высокой волны
в этих случаях оказалось чуть менее отчетливым.

Наконец, для полноты картины кратко упомянем еще одну серию численных
экспериментов. Поскольку при частоте в середине щели отражение от массива
барьеров достаточно резкое, можно надолго ``запирать'' волну в относительно
неширокой области, где высота барьеров уменьшается до нуля (а в остальных
местах барьеры присутствуют). Моделирование показало, что большая часть
энергии волны остается запертой на протяжении сотен периодов.

{\bf Заключение.}
Таким образом, квазимонохроматическая волна  на воде при взаимодействии
с плавно неоднородным массивом донных препятствий способна сформировать сильно
нелинейную локализованную структуру. Явление это имеет резонансный характер
и возможно лишь в узком диапазоне частот внутри запрещенной зоны.

Надо отметить, что получить настоящие долгоживущие Брэгговские солитоны
таким способом не удалось. Высокая волновая группа отражается в обратном
направлении, а не проникает далее в барьерную область. Поэтому необходимо
продолжить исследование данной проблемы.

{\bf Финансирование работы.}
Работа выполнена в рамках госзадания по теме FFWR-2024-0013.

{\bf Конфликт интересов.}
Автор данной работы заявляет, что у него нет конфликта интересов.

\end{document}